\documentclass[prd,aps,showpacs,onecolumn]{revtex4}
\usepackage{amsfonts}
\usepackage{amsmath}
\usepackage{graphicx}
\usepackage{bm}

\begin{document}

\title{Duality properties of Gorringe-Leach equations}
\author{Yves Grandati, Alain B\'erard and Herv\'e Mohrbach}
\affiliation{Laboratoire de Physique Mol\'eculaire et des
Collisions,
 Institut de
Physique, ICPMB, IF CNRS 2843, Universit\'e Paul Verlaine-Metz, Bd
 Arago,
57078 Metz, Cedex 3, France}

\begin{abstract}
In the category of motions preserving the angular momentum's
direction, Gorringe and Leach exhibited two classes of
differential equations
 having
elliptical orbits. After enlarging slightly these classes, we show
that they are related by a duality correspondence of the
Arnold-Vassiliev
 type.
The specific associated conserved quantities (Laplace-Runge-Lenz
vector
 and
Fradkin-Jauch-Hill tensor) are then dual reflections one of the
other.
\end{abstract}

\maketitle


\section{Introduction}

In 1993, Gorringe and Leach \cite{GL1}, exhibited two classes of
differential equations incorporating drag terms which have closed
 elliptical
orbits, generalizing then previous results of Jezewski and
Mittleman
 \cite%
{JM,JM2} and Leach \cite{L}. Both possess conserved quantities
which
 extend
the Laplace-Runge-Lenz vector and Fradkin-Jauch-Hill tensor
 respectively.
These two classes belong to a broader category of planar motions
incorporating velocity dependent terms submitted to certain
constraint
 on
their coefficients \cite{GL2,GL3,GL4,LF1}.

In this paper we show that the above classes can be slightly
enlarged.
 To
these generalized Gorringe-Leach equations are associated two
conserved quantities : a pseudo-energy and a pseudo-angular
momentum. We obtain
 two
types (H and K) of generalized Gorringe-Leach equations presenting
 closed
orbits for every values of these quantities. In the spherically
 symmetrical
case, we obtain compact analytical formulas for the periods which,
if restricted to the standard case, recover the results of ref.
\cite{GL1}.
 The
H and K types belongs to a larger category of generalized
 Gorringe-Leach
equations possessing duality properties (in the Arnold-Vassiliev
sense
 \cite%
{AV,GBMo}). In this category, the equations can be gathered in
classes indexed by a characteristic real parameter $\nu $. Each
$\nu $-class possesses an associated dual $\mu $-class, with $\mu
=-\frac{\nu }{1+\frac{\nu }{2}}$. H type and K type generalized
Gorringe-Leach equations are then shown to be dual of each other.
As in the conservative case \cite{GBMo}, the pseudo
Laplace-Runge-Lenz vector associated to the K type equations is
also proportional to the dual transform of the pseudo
Fradkin-Jauch-Hill
 tensor
of the H type equations.

\section{Complex description of motions with conservation of the
 direction
of angular momentum}

As was shown in \cite{LF1}, the most general form for the equation
of a 3-dimensional motion $\overrightarrow{r}(t)$ for which the
angular
 momentum $%
\overrightarrow{L}=L\overrightarrow{e}_{L}$ conserves its
direction is
 given
by
\begin{equation}
\overset{..}{\overrightarrow{r}}+h\overset{.}{\overrightarrow{r}}+g%
\overrightarrow{r}=0,
\end{equation}
where $h$ and $g$ are two arbitrary scalars depending on time $t$.

As the motion is confined to the plane orthogonal to
 $\overrightarrow{e}_{L}$%
, we can adopt a complex representation for the position
 $\overrightarrow{r}%
(t)\rightarrow z(t)$.

The functions $h$ and $g$ are then represented by two
\textbf{real} arbitrary functions of $t$ through $z(t)$ and
$\overset{.}{z}(t)$, not necessarily analytical (that is
potentially dependent on
 $\overline{z}(t)$
and $\overset{.}{\overline{z}}(t)$).

We apply our interest more specifically to the autonomous case
($g$ and
 $h$
do not depend explicitly on $t$), to which we add the constraint
that
 the
last term is a function of $z$ and $\overline{z}$ only. We then
arrive
 to
the following equation for our planar motion $z(t)$ :
\begin{equation}
\overset{..}{z}+h(z,\overline{z},\overset{.}{z},\overset{.}{\overline{z}})%
\overset{.}{z}+g(z,\overline{z})z=0  \label{2}
\end{equation}%
$g$ and $h$ being two arbitrary real-valued functions.

\section{Euler-Sundman reparametrization}

We now perform an Euler-Sundman reparametrization $t\in \mathbb{R}%
^{+}\rightarrow s\in \mathbb{R}^{+}$ (where the correspondence is
one
 to one
and increasing) of our motion, $z(t)$. We put
\begin{equation}
s=s(t,z(t),\overline{z}(t))
\end{equation}%
with
\begin{equation}
\frac{ds}{dt}=\overset{.}{s}\ >0.
\end{equation}%
If we write $\frac{df}{ds}=f^{\prime }$, Eq. \ref{2} becomes
\begin{equation}
z^{\prime \prime
 }+\frac{\overset{..}{s}+\overset{.}{s}h(z,\overline{z},%
\overset{.}{s}z^{\prime },\overset{.}{s}\overline{z}^{\prime
})}{\left( \overset{.}{s}\right) ^{2}}z^{\prime
}+\frac{g(z,\overline{z})}{\left( \overset{.}{s}\right) ^{2}}z=0
\label{5}
\end{equation}%
in which $s$ is called the \textbf{pseudotime} and $z(s)$ the \textbf{%
pseudomotion}.

If we choose the reparametrization in such a way that
\begin{equation}
\overset{..}{s}+\overset{.}{s}h(z,\overline{z},\overset{.}{s}z^{\prime
 },%
\overset{.}{s}\overline{z}^{\prime })=0,
\end{equation}%
then Eq. \ref{5} for the pseudomotion becomes
\begin{equation}
z^{\prime \prime }+g(z,\overline{z})\left( e^{2\int
 h(z,\overline{z},\overset%
{.}{z},\overset{.}{\overline{z}})dt}\right) _{t=t(s)}z=0.
\label{7}
\end{equation}

The problem is considerably simplified if we restrict $h$ to be a
total derivative :
\begin{equation}
h(z,\overline{z},\overset{.}{z},\overset{.}{\overline{z}})=\overset{.}{H}(z,%
\overline{z})
\end{equation}%
In this particular case the initial Eq.\ref{2} is written as
\begin{equation}
\overset{..}{z}+\overset{.}{H}(z,\overline{z})\overset{.}{z}+g(z,\overline{z}%
)z=0  \label{9}
\end{equation}%
The reparametrization is given by :
\begin{equation}
ds\ =e^{-H(z,\overline{z})}dt
\end{equation}%
and the pseudomotion Eq. \ref{7}, takes the following simple form
\begin{equation}
z^{\prime \prime }+g(z,\overline{z})e^{2H(z,\overline{z})}z=0
\end{equation}%
which is the equation of an autonomous conservative motion. The
term
 $g(z,%
\overline{z})e^{2H(z,\overline{z})}z$ is called the
 \textbf{pseudoforce}.

From now we always place ourselves in this case.

\section{Radial pseudoforce}

Consider the case in which the pseudoforce, $g(z,\overline{z})e^{2H(z,%
\overline{z})}z,$ is derived from a radial real-valued potential
 $U(r)$:
\begin{equation}
\overrightarrow{\nabla }U(\overrightarrow{r})=2\frac{\partial
 U(r)}{\partial
\overline{z}}=\varphi
(r)z=g(z,\overline{z})e^{2H(z,\overline{z})}z
\end{equation}%
where
\begin{equation}
\varphi (r)=\frac{1}{r}\frac{\partial U(r)}{\partial r}.
\label{13}
\end{equation}

The initial motion Eq. \ref{9}, takes the form:
\begin{equation}
\overset{..}{z}+\overset{.}{H}(z,\overline{z})\overset{.}{z}+\varphi
(r)e^{-2H(z,\overline{z})}z=0  \label{14}
\end{equation}%
which we call the \textbf{generalized Gorringe-Leach equation }
 \cite{GL1}.

To the pseudomotion we can associate a \textbf{pseudo-angular
 momentum},
\begin{equation}
\mathbf{\ }\overrightarrow{\mathcal{L}}=\overrightarrow{r}\times
\overrightarrow{r}^{\prime }\equiv \frac{1}{2i}\left( z^{\prime
 }\overline{z}%
-z\overline{z}^{\prime }\right) \overrightarrow{e}_{L},
\label{15}
\end{equation}%
and a \textbf{pseudoenergy},
\begin{equation}
\mathbf{\ }\mathcal{E}=\frac{1}{2}\left\vert
\overrightarrow{r}^{\prime }\right\vert ^{2}+U(r)\equiv
\frac{1}{2}\left\vert z^{\prime
 }\right\vert
^{2}+U(r)),  \label{16}
\end{equation}%
which are constants of the pseudomotion.

Consquently the original motion possesses also the two conserved
 quantities:
\begin{equation}
\left\{
\begin{array}{c}
\overrightarrow{\mathcal{L}}=\mathcal{L}\overrightarrow{e}_{L},\quad
\mathcal{L}=e^{H(z,\overline{z})}L \\
\mathcal{E}=e^{2H(z,\overline{z})}\frac{\left|
\overset{.}{z}\right|
 ^{2}}{2}%
+U(r).%
\end{array}
\right.
\end{equation}

\section{Bertrand's theorem}

Applying Bertrand's theorem \cite{Bert,GBM} to the pseudomotion,
we
 deduce
immediately that the only pseudomotions of which orbits are closed
for
 every
value of the characteristic parameters (given by the pseudo-energy
and
 the
pseudo-angular momentum), are those associated to the Hooke and
Kepler potentials,
\begin{equation}
U_{H}(r)=\frac{1}{2}kr^{2}
\end{equation}
and
\begin{equation}
U_{K}(r)=-\frac{k}{r}
\end{equation}
In both cases the orbits are ellipses, centered in O in the H case
and having a focus in O in the K case.

The orbits of the pseudomotion being also those of the initial
motion
 (the
two motions differing only by a reparametrization), we deduce that
the
 only
generalized Gorringe-Leach Eq.\ref{14}, which always (that is for
every value of the conserved pseudo-energy $\mathcal{E}$ and
pseudo-angular
momentum $\mathcal{L}$) admit closed orbits, are those for which $g(z,%
\overline{z})$ is of the form:
\begin{equation}
g_{H}(z,\overline{z})=ke^{-2H(z,\overline{z})}  \label{20}
\end{equation}%
or
\begin{equation}
g_{K}(z,\overline{z})=\frac{k}{r^{3}}e^{-2H(z,\overline{z})}.
  \label{21}
\end{equation}%
In other words, the initial equation must be of the form
\begin{equation}
\overset{..}{z}+\overset{.}{\widehat{H(z,\overline{z})}}\overset{.}{z}%
+ke^{-2H(z,\overline{z})}z=0,\quad \text{type H equation,}
\label{22}
\end{equation}%
or:
\begin{equation}
\overset{..}{z}+\overset{.}{\widehat{H(z,\overline{z})}}\overset{.}{z}%
+kr^{-3}e^{-2H(z,\overline{z})}z=0,\quad \text{type K equation.}
  \label{23}
\end{equation}

\section{Spherically symmetrical case}

In the spherically symmetrical case $H$ and $g$ depend only on the
 radial
variable, $r=\sqrt{z\overline{z}}$, ie
\begin{equation}
\left\{
\begin{array}{c}
H(z,\overline{z})=H(z\overline{z})=\alpha (r) \\
g(z,\overline{z})=g(z\overline{z})=u(r),%
\end{array}%
\right.
\end{equation}%
and we are in the case in which the pseudoforce derives from a
radial potential given by
\begin{equation}
U(r)=\int ru(r)e^{2\alpha (r)}dr.
\end{equation}%
In this case the initial Eq.\ref{14} is written as
\begin{equation}
\overset{..}{z}+\overset{.}{\alpha }(r)\overset{.}{z}+u(r)z=0.
  \label{26}
\end{equation}%
The two conserved quantities Eqs. \ref{17}  become:
\begin{equation}
\left\{
\begin{array}{c}
\overrightarrow{\mathcal{L}}=\mathcal{L}\overrightarrow{e}_{L},\quad
\mathcal{L}=e^{\alpha (r)}L \\
\mathcal{E}=e^{2\alpha (r)}\frac{\left\vert
\overset{.}{z}\right\vert
 ^{2}}{2%
}+U(r)%
\end{array}%
\right.
\end{equation}%
and the type H, and K Eqs. \ref{22},\ref{23} take the forms
\begin{equation}
\left\{
\begin{array}{c}
\overset{..}{z}+\overset{.}{\alpha }(r)\overset{.}{z}+ke^{-2\alpha
(r)}z=0,\quad \text{type H equation, } \\
\overset{..}{z}+\overset{.}{\alpha
 }(r)\overset{.}{z}+kr^{-3}e^{-2\alpha
(r)}z=0,\quad \text{type K equation. }%
\end{array}%
\right.
\end{equation}

Note that, if we restrict ourselves to the case of a logarithmic
 function
for $\alpha (r)$, we recover here in a very direct manner the
original Gorringe-Leach results \cite{GL1,LF1}.

\section{Orbital period in the spherical symmetrical case}

The radial component $r(s)$ of the pseudomotion is given by
integration
 of
Barrow's differential formula applied to the radial pseudomotion
 \cite{GBM},
which gives
\begin{equation}
dt=\frac{ds}{\overset{.}{s}}=\frac{1}{\sqrt{2}}\frac{e^{\alpha
 (r)}dr}{\sqrt{%
\mathcal{E}-V_{\mathcal{L}}(r)}},
\end{equation}%
where the radial effective potential $V_{L^{\prime }}(r)$ is
\begin{equation}
V_{\mathcal{L}}(r)=U(r)+\frac{\mathcal{L}^{2}}{2r^{2}}.
\label{30}
\end{equation}

When the orbit is bounded, the radial variable $r$ oscillates
between
 the
extremal values $a$ et $b$ (pericentral and apocentral radii)
which are roots of the numerical equation:
\begin{equation}
\mathcal{E}-V_{\mathcal{L}}(r)=0.
\end{equation}

The orbital period is
\begin{equation}
T=\sqrt{2}\int_{r_{<}}^{r_{>}}\frac{e^{\alpha
 (r)}dr}{\sqrt{\mathcal{E}-V_{%
\mathcal{L}}(r)}}.
\end{equation}
We are always in this case for the type H and K equations.

If, for the corresponding elliptical orbits, we note $A$ the minor
axis
 and $%
B$ the major axis, we have :

* In the H case ($\mathcal{E}>0$)
 $V_{\mathcal{L}}^{H}(r)=\frac{1}{2}kr^{2}+%
\frac{\mathcal{L}^{2}}{2r^{2}}$,
\begin{equation}
\left\{
\begin{array}{c}
A=a \\
B=b%
\end{array}
\right.
\end{equation}
and
\begin{equation}
T_{H}=\frac{2}{\sqrt{k}}\int_{a}^{b}\frac{re^{\alpha
 (r)}dr}{\sqrt{\left(
r^{2}-a^{2}\right) \left( b^{2}-r^{2}\right) }}.
\end{equation}

* In the K case ($\mathcal{E}<0$)
 $V_{\mathcal{L}}^{K}(r)=-\frac{k}{r}+\frac{%
\mathcal{L}^{2}}{2r^{2}}$,
\begin{equation}
\left\{
\begin{array}{c}
b=B+\sqrt{B^{2}-A^{2}} \\
a=B-\sqrt{B^{2}-A^{2}}%
\end{array}%
\right.  \label{35}
\end{equation}%
and

\begin{equation}
T_{K}=\frac{\sqrt{2}}{\sqrt{-\mathcal{E}}}\int_{a}^{b}\frac{re^{\alpha
 (r)}dr%
}{\sqrt{\left( r-a\right) \left( b-r\right) }}.
\end{equation}

By a straightforward changes of variables we finally obtain :
\begin{equation}
\left\{
\begin{array}{c}
T_{H}=\frac{1}{\sqrt{k}}\int_{0}^{1}x^{-\frac{1}{2}}\left(
1-x\right)
 ^{-%
\frac{1}{2}}\exp \left( \alpha (\sqrt{\left( b^{2}-a^{2}\right)
 x+a^{2}}%
)\right) dx \\
T_{K}=\frac{\sqrt{2}\left( b-a\right)
 }{\sqrt{-\mathcal{E}}}\int_{0}^{1}x^{-%
\frac{1}{2}}\left( 1-x\right) ^{-\frac{1}{2}}(x+\frac{a}{b-a})\exp
 \left(
\alpha (\left( b-a\right) x+a)\right) dx%
\end{array}%
\right.  \label{37}
\end{equation}

\section{Gorringe-Leach equations}

Gorringe-Leach equations \cite{GL1} are equations of the type
 considered with

\begin{equation}
\alpha (r)=-\frac{\alpha }{2}\ln (r).
\end{equation}

The initial equations of motion presenting closed orbits
 are
then (see Eq. \ref{28}) :
\begin{equation}
\overset{..}{z}-\frac{\alpha
 }{2}\frac{\overset{.}{r}}{r}\overset{.}{z}%
+kr^{\alpha }z=0,\quad \text{ H type,}
\end{equation}%
and
\begin{equation}
\overset{..}{z}-\frac{\alpha
 }{2}\frac{\overset{.}{r}}{r}\overset{.}{z}%
+kr^{\alpha -3}z=0,\quad \text{ K type.}  \label{40}
\end{equation}

As for the corresponding orbital periods Eq. \ref{37}, they are
written
 in
these cases as
\begin{equation}
T_{H}=\frac{\pi a^{-\frac{\alpha }{2}}}{\sqrt{k}}F\left(
\frac{\alpha
 }{4},%
\frac{1}{2},1,\frac{a^{2}-b^{2}}{a^{2}}\right)
\end{equation}%
and
\begin{equation}
T_{K}=\frac{\sqrt{2}\pi a^{1-\frac{\alpha }{2}}}{\sqrt{-E^{\prime
 }}}F\left(
\frac{\alpha }{2}-1,\frac{1}{2},1,\frac{a-b}{a}\right) ,
\end{equation}%
where $F$ is an hypergeometric function.

Note that the hypergeometric functions of the form $F\left( \nu
,\beta ,2\beta ,z\right) $ can be rewritten in terms of Legendre
functions via
 :
\begin{equation}
F\left( \nu ,\beta ,2\beta ,z\right) =2^{2\beta -1}\Gamma (\beta
 +\frac{1}{2}%
)z^{\frac{1}{2}-\beta }(1-z)^{\frac{\beta -\nu
-\frac{1}{2}}{2}}P_{\nu -\beta -\frac{1}{2}}^{\frac{1}{2}-\beta
}\left(
 \frac{1-\frac{z}{2}}{\sqrt{%
1-z}}\right) .
\end{equation}%
Under this form we recover the results of Gorringe and Leach
 \cite{GL1},
namely
\begin{equation}
T_{H}=\frac{\pi a^{-\frac{\alpha
}{2}}}{\sqrt{k}}(1-z)^{-\frac{\alpha
 }{8}%
}P_{\frac{\alpha }{4}-1}\left(
\frac{1-\frac{z}{2}}{\sqrt{1-z}}\right)
\end{equation}%
and
\begin{equation}
T_{K}=\frac{\sqrt{2}\pi a^{1-\frac{\alpha }{2}}}{\sqrt{-E^{\prime
 }}}(1-z)^{-%
\frac{\alpha -2}{2}}P_{\frac{\alpha }{2}-2}\left(
 \frac{1-\frac{z}{2}}{\sqrt{%
1-z}}\right) .
\end{equation}

\section{Arnold-Vassiliev duality}

In order for the pseudomotion potential be an Arnold-Vassiliev
 potential
\cite{AV,GBMo} and then be dualizable in the Arnold-Vassiliev
sense, it
 has
to be of the type
\begin{equation}
U(z,\overline{z})=k\left\vert u(z)\right\vert ^{2}\in \mathbb{R}
  \label{46}
\end{equation}%
It is easy to show that, if we suppose $u(z)$ analytical, this
 necessitates
\begin{equation}
z\frac{u^{\left( 1\right) }(z)}{u(z)}=cste=\frac{\nu }{2}\in
 \mathbb{R},
\end{equation}%
that is, $U$ must be a central power law potential namely
\begin{equation}
U(z,\overline{z})=Ar^{\nu }
\end{equation}%
and the pseudomotion equation is ($k=\nu A$)
\begin{equation}
z^{\prime \prime }+kr^{\nu -2}z=0.  \label{49}
\end{equation}

In other words the generalized Gorringe-Leach equations (Eq.
\ref{14})
 which
are dualizable in the Arnold-Vassiliev sense have the form
\begin{equation}
\overset{..}{z}+\overset{.}{H}(z,\overline{z})\overset{.}{z}+kr^{\nu
-2}e^{-2H(z,\overline{z})}z=0.  \label{50}
\end{equation}%
Such an equation is said to be of class $\nu $. In this
nomenclature an H-type equation is then an equation of class 2 and
a K-type equation an equation of class -1.

The pseudomotion corresponding to an equation of class $\nu $
admits a
 dual
in the Arnold-Vassiliev sense, the associated potential of which
is
 given by
\begin{equation}
V(w,\overline{w})=B\rho ^{\mu },  \label{51}
\end{equation}%
where $\rho =\left\vert w\right\vert ,\ B=-\mathcal{E}/\left(
 1+\frac{\mu }{2%
}\right) ^{2},\ \left( 1+\frac{\mu }{2}\right) \left( 1+\frac{\nu }{2}%
\right) =1$.

The dual pseudomotion equation is then

\begin{equation}
w^{\prime \prime }+\kappa r^{\mu -2}w=0
\end{equation}
with $\kappa =\mu B=-\mu E^{\prime }/\left( 1+\frac{\mu
}{2}\right)
 ^{2}$.

The correspondence between the position variables is
\begin{equation}
w=\left( 1+\frac{\mu }{2}\right) z^{1+\frac{\mu }{2}}  \label{52}
\end{equation}%
and the dual pseudotimes are related by the Euler-Sundman
 reparametrization
\begin{equation}
ds=\left( 1+\frac{\nu }{2}\right) ^{-\frac{\nu }{1+\frac{\nu
}{2}}}\rho
 ^{-%
\frac{\nu }{1+\frac{\nu }{2}}}d\sigma .  \label{53}
\end{equation}

To the pseudo-motion, $w(\sigma ),$ is also associated a
 \textquotedblright
true\textquotedblright\ motion, $w(\tau )$, satisfying a
generalized Gorringe-Leach equation of the form
\begin{equation}
\overset{..}{w}+\overset{.}{\widetilde{H}}(z,\overline{z})\overset{.}{w}%
+\kappa r^{\mu -2}e^{-2\widetilde{H}(z,\overline{z})}w=0,
\label{55}
\end{equation}%
$\widetilde{H}(z,\overline{z})$ being an arbitrary real valued
 function.

The dot indicates here the derivative with respect to $\tau $,
where
 the
\textquotedblright true\textquotedblright\ time, $\tau ,$ is
related to
 the
pseudotime, $\sigma ,$ by the Euler-Sundman reparametrization
\begin{equation}
d\sigma \ =e^{-\widetilde{H}(z,\overline{z})}d\tau .
\end{equation}%
This equation is an equation of class $\mu $. We can then see that
the generalized Gorringe-Leach equations, if not strictly
 dualizable,
can, however, be grouped into classes indexed by a real
characteristic exponent, the $\nu $ class being in dual
correspondence with the class
 $\mu $%
, such that :
\begin{equation}
\left( 1+\frac{\mu }{2}\right) \left( 1+\frac{\nu }{2}\right) =1.
\end{equation}

The classes $2$ and $-1$ are dual in the sense that they are
linked by
 a
Levi-Civita \cite{GBMo} change of coordinates, this preserving the
 elliptical
structure of the orbits. The fact that the equations of $H$ and
$K$
 types
possess the same orbital characteristics is not fortuitous, but is
a
 direct
consequence of this duality.

As was demonstrated in \cite{GBMo}, in complex formulation the
 existence of
an additional conserved quantity for a motion in the Hooke
potential
 (the
Fradkin-Jauch-Hill tensor) is evident. If we apply this result to
a
 type H
pseudomotion, we deduce immediately that a H type generalized
 Gorringe-Leach
equation admits besides its pseudo-energy, $\mathcal{E}$, an
additional conserved complex quantity, namely

\begin{equation}
\mathcal{T}=\frac{1}{2}\left( z^{\prime }\right) ^{2}+\frac{1}{2}%
kz^{2}=e^{2H(z,\overline{z})}\frac{1}{2}\left(
\overset{.}{z}\right)
 ^{2}+%
\frac{1}{2}kz^{2}.
\end{equation}%
(The components of the 2D associated pseudo-FJH tensor are given
by
 $T_{xx}=%
\text{Re}\left( \mathcal{E}+\mathcal{T}\right) $,
 $T_{yy}=\text{Re}\left(
\mathcal{E}-\mathcal{T}\right) $ and $T_{xy}=\text{Im}\left(
 \mathcal{T}%
\right) =T_{yx}.$)

Following \cite{GBMo} the dual K-type generalized Gorringe-Leach
 equation $%
\overset{..}{w}+\overset{.}{\widehat{\widetilde{H}(w,\overline{w})}}\overset{%
.}{w}+\widetilde{k}\rho
^{-3}e^{-2\widetilde{H}(w,\overline{w})}w=0,$
 admits
in addition to its pseudo-energy $\widetilde{\mathcal{E}}$, the
 following
conserved complex quantity :

\begin{equation}
\mathcal{A}=\frac{\mathcal{T}}{2\widetilde{k}}=\frac{1}{\widetilde{k}}%
iw^{\prime }\widetilde{\mathcal{L}}-\frac{w}{\rho
 }=e^{2\widetilde{H}(w,%
\overline{w})}\frac{m}{\widetilde{k}}iw^{\prime
 }\widetilde{L}-\frac{w}{\rho
}  \label{59}
\end{equation}%
which is the image of $\mathcal{T}$ via the dual transformation
 (Eqs.\ref{52}
and \ref{53}), with $\widetilde{\mathcal{E}}=-\frac{k}{2}$ and
 $\widetilde{k}%
=-\frac{\mathcal{E}}{2}$. We clearly recognize in $\mathcal{A}$
the
 complex
formulation of the pseudo-Laplace-Runge-Lenz vector \cite{GL1} :

\begin{equation}
\overrightarrow{A}=\frac{e^{2\widetilde{H}}}{\widetilde{k}}\overrightarrow{%
\widetilde{L}}\times \overrightarrow{\rho }^{\prime
 }-\frac{\overrightarrow{%
\rho }}{\rho }.
\end{equation}

\section{Acknowledgments}

We would like to thank Professor P.G.L. Leach for useful
suggestions
 and a
careful reading of the manuscript.

\end{document}